\documentclass[prb,twocolumn,amsmath,superscriptaddress]{revtex4}
\usepackage{graphicx,color}
\usepackage{amssymb}
\newcommand{\He}{$^3$He}
\newcommand{\Hefour}{$^4$He}
\newcommand{\be}{\begin{equation}}
\newcommand{\ee}{\end{equation}}
\newcommand{\ber}{\begin{eqnarray}}
\newcommand{\eer}{\end{eqnarray}}
\def\vm{{\bf m}}
\def\vn{{\bf n}}

\def\vB{{\bf B}}
\def\vS{{\bf S}}
\def\vf{{\bf f}}
\def\vp{{\bf p}}

\def\vz{{\bf z}}
\newcommand{\vSigma}{\mbox{\boldmath$\Sigma$}}
\newcommand{\vSig}{{\mbox{\footnotesize $\vSigma$}}}
\newcommand{\vsigma}{\mbox{\boldmath$\sigma$}}
\newcommand{\Sig}{{\mbox{\footnotesize $\Sigma$}}}
\newcommand{\hSig}{{\hat{\mbox{\footnotesize $\Sigma$}}}}
\newcommand{\Del}{{\mbox{\footnotesize $\Delta$}}}
\newcommand{\hDel}{{\hat{\mbox{\footnotesize $\Delta$}}}}

\newcommand{\vDelta}{\mbox{\boldmath$\Delta$}}
\newcommand{\htm}{\hat{\mathfrak{t}}}
\newcommand{\htu}{\hat{\mathfrak{u}}}
\newcommand{\hg}{\hat{\mathfrak{g}}}

\newcommand{\ess}{\textsf{s}}

\newcommand{\x}{\tilde{x}}
\newcommand{\onehalf}{\frac{\text{\footnotesize 1}}{\text{\footnotesize 2}}}
\newcommand{\onethird}{\frac{\mbox{\small 1}}{\mbox{\small 3}}}
\newcommand{\onematrix}{\mbox{${\mathfrak{1}}$}}
\newcommand{\tinyonehalf}{\frac{\mbox{\tiny 1}}{\mbox{\tiny 2}}}
\newcommand{\tinyonethird}{\frac{\mbox{\tiny 1}}{\mbox{\tiny 3}}}
\def\sgn#1{\text{sgn}(#1)}

\begin{document}
\preprint{cond-mat/2003}
%\title{Effect of Magnetic Scattering on the Phase Diagram of \He\ in Aerogel}
%\title{Possible Evidence of Magnetic Scattering from the Phase Diagram of \He\ in Aerogel}
\title{Impurity Effects on the A$_1$-A$_2$ Splitting of Superfluid \He\ in Aerogel}
\author{J. A. Sauls}%\email{sauls@snowmass.phys.nwu.edu}
\affiliation{Department of Physics and Astronomy, Northwestern University, Evanston, IL, USA}
\affiliation{Centre de Recherches sur les Tr\`es Basses Temp\'eratures,
             Centre National de la Recherche Scientifique,\\
             Laboratoire Associ\'e \`a l'Universit\'e J. Fourier,
            BP 166, 38042 Grenoble Cedex 9, France}
\author{Priya Sharma}
\affiliation{Department of Physics and Astronomy, Northwestern University, Evanston, IL, USA}
\date{\today}
%~~~~~~~~~~~~~~~~~~~~~~~~~~~~~~~~~~~~~~~~~~~~~~~~~~~~~~~~~~~~~~~~~~~~~~~~~~~~~~~~~~~~~~~~~~~
\begin{abstract}
When liquid \He\ is impregnated into silica aerogel a solid-like layer of \He\ atoms
coats the silica structure. The surface \He\ is in fast exchange with the liquid on NMR
timescales. The exchange coupling of liquid \He\ quasiparticles with the localized \He\
spins modifies the scattering of \He\ quasiparticles by the aerogel structure. In a
magnetic field the polarization of the solid spins gives rise to a splitting of the
scattering cross-section of for `up' vs. `down' spin quasiparticles, relative to the
polarization of the solid \He. We discuss this effect, as well as the effects of
non-magnetic scattering, in the context of a possible
splitting of the superfluid transition for $\uparrow\uparrow$ vs. $\downarrow\downarrow$
Cooper pairs for superfluid \He\ in aerogel, analogous to the $A_1-A_2$ splitting in bulk \He.
Comparison with the existing measurements of $T_c$ for $B< 5\,\text{kG}$, which show no
evidence of an $A_1-A_2$ splitting, suggests
a liquid-solid exchange coupling of order $J\simeq 0.1\,\text{mK}$.
Measurements at higher fields, $B \gtrsim 20\text{kG}$, should saturate
the polarization of the solid \He\ and reveal the $A_1-A_2$ splitting.
\end{abstract}
%~~~~~~~~~~~~~~~~~~~~~~~~~~~~~~~~~~~~~~~~~~~~~~~~~~~~~~~~~~~~~~~~~~~~~~~~~~~~~~~~~~~~~~~~~~~
\maketitle
%~~~~~~~~~~~~~~~~~~~~~~~~~~~~~~~~~~~~~~~~~~~~~~~~~~~~~~~~~~~~~~~~~~~~~~~~~~~~~~~~~~~~~~~~~~~

%\section*{\label{sec:level1}Introduction}

One focus of experimental investigations of \He\ in aerogel has been the
determination of the phase diagram. Torsional oscillator, NMR, vibrating wire and
sound attenuation experiments on \He\ in $\approx 98\%$ porosity aerogels,
suggest that there is just one superfluid phase in zero magnetic field over the
pressure range, $34\,\text{atm}<p\lesssim 6\,\text{atm}$.\cite{bar00,bru01,ger01}
At lower pressures there appears to be only normal \He\ down to zero
temperature.\cite{mat97} In short, a B-like phase, with reduced susceptibility,
is the only stable superfluid
phase observed in zero field. An A-like phase, if it exists in zero field, may be
stable only in a very small region $\Del T\le 20\mu\mbox{K}$ near $T_c$. As in bulk
\He\, an A-like phase can be stabilized in a magnetic field; the region
of stability, $T_{\text{\tiny AB}}\le T < T_c$, increases quadratically with field,
$T_c-T_{\text{\tiny AB}}= -g_{\text{\tiny AB}}\,B^2$. However, a splitting of the
transition for $\uparrow\uparrow$ and $\downarrow\downarrow$ pairs, analogous to
the $A_1-A_2$ splitting in bulk \He, has so far not been observed for
fields up to $B\approx 5\,\text{kG}$.\cite{ger01,ger02}

In pure \He, on application of a magnetic field, the $A_1$ phase, characterized by
spin-polarized Cooper pairs composed of only $\uparrow\uparrow$ spins nucleates at a
temperature slightly higher than the zero-field transition,
$T_c^{A_1}=T_c+\lambda^{\text{\tiny A}_1}\,B$.\footnote{The $A_1$ phase
corresponds to pairs of \He\ quasiparticles with their magnetic moments aligned
along the field. We follow the notation of Ref. \onlinecite{vollhardt90} and
define the quantization axis for the spin to be $\hat{\vz}||-\vB$ to compensate
for the negative gyromagnetic ratio of \He.}
This transition is followed by a second
transition, shifted below the zero-field transition, at $T_c^{A_2}=T_c-\lambda^{\text{\tiny
A}_2}\,B$, in which the $\downarrow\downarrow$ pairs nucleate. The region of stability
of \textsl{pure} $\uparrow\uparrow$ pairs increases linearly with field, $\Del
T_{A_1-A_2}\equiv(T_c^{A_1}-T_c^{A_2})= \left(\lambda^{\text{\tiny
A}_1}+\lambda^{\text{\tiny A}_2}\right) B$, at a rate of $\lambda^{\text{\tiny
A}_1}+\lambda^{\text{\tiny A}_2}\simeq 6.1\mu\mbox{K/kG}$
at $p=33.4\,\text{bar}$.\cite{isr84,sag84}

This splitting of the zero-field transition originates from the
combined effect of the nuclear Zeeman coupling to the \He\ spin,
and particle-hole asymmetry in the normal-state density of states and
pairing interaction. The original estimate of $\lambda^{\text{\tiny A}_1}$
by Ambegaokar and Mermin\cite{amb73} was based on the asymmetry of the
density of states for $\uparrow$ vs. $\downarrow$ quasiparticles at the
Fermi level. More involved calculations include the effects of
spin-polarization of the Fermi liquid on the pairing
interaction.\cite{lev81,bed84} Estimates of the splitting of
the $A_1$ transition are of order
%~~~~~~~~~~~~~~~~~~~~~~~~~~~~~~~~~~~~~~~~~~~~~~~~~~~~~~~~~~~~~~~~~~~~~~~~~
\be
\lambda^{\text{\tiny A}_1}=\Lambda\,\Big\vert\frac{\gamma\hbar}{2}\Big\vert
                           \left(\frac{k_BT_{c}}{E_f}\right)
\,,
\ee
%~~~~~~~~~~~~~~~~~~~~~~~~~~~~~~~~~~~~~~~~~~~~~~~~~~~~~~~~~~~~~~~~~~~~~~~~~~
where $\gamma$ is the gyromagnetic ratio for \He, $k_B T_c/E_f\sim
10^{-3}$ and $\Lambda\sim{\cal O}(1-10)$ is a ``high-energy'' vertex. A
first principles calculation of $\Lambda$ requires a solution of the
many-body problem for the pairing interaction in \He. Alternatively, we
treat $\Lambda$ on the same level as other high-energy vertices in
Fermi liquid theory; $\Lambda$ is a Fermi-liquid parameter, which can be
determined by comparing physical predictions of the Fermi liquid theory
with experiment - in this case the $A_1-A$ splitting. Once determined,
the effects of $\Lambda$ on other low-energy properties of \He\ can be
calculated. Thus, the $A_2$ transition can be calculated in terms of
$\Lambda$, and corrections to $T_c^{A_2}$ from the
$\uparrow\uparrow$ pair condensate.\cite{mer73}

The disorder introduced by the aerogel structure into liquid \He\
is, on average, weak on the high-energy scale, $\hbar/\tau E_f\ll
1$. Thus, the Fermi-liquid interactions are essentially unaffected
by the aerogel, and we can calculate the effects of aerogel on the
low-energy excitations and superfluid properties within Fermi liquid
theory. The main effect of the aerogel structure is to scatter \He\
quasiparticles moving with the Fermi velocity. At temperatures below
$T^*\approx 10\,\text{mK}$ elastic scattering by the aerogel
dominates inelastic quasiparticle-quasiparticle
collisions.\cite{rai98} This limits the mean free path of normal
\He\ quasiparticles to $\ell\simeq 130-180\,\text{nm}$. In p-wave
superfluids quasiparticle scattering is intrinsically pairbreaking
and leads to renormalization of nearly all properties of the
superfluid phases, including the $A_1$ and $A_2$ transition
temperatures. The suppression of $T_c$, as well as pair-breaking
effects on observable properties of the superfluid phases, have been
analyzed theoretically for non-magnetic
scattering.\cite{thu98,thu98c,bar96,bar01,sha01,min02}
Here we analyze the effects of scattering by the aerogel
on the $A_1$ and $A_2$ transitions.

The aerogel has a strong effect on the \textsl{short-distance}, high-energy
properties of the liquid locally near the silica strands. The first few
atomic layers of \He\ are adsorbed on the silica structure and form a highly
polarizable solid-like phase, observable as a Curie-like component
of the magnetization of \He-aerogel.\cite{spr95} The surface \He\ is in fast
exchange with the liquid on typical NMR timescales, implying a liquid-solid
exchange interaction, $|J|/h\gtrsim 0.66\,\text{MHz}$ ($|J|\gtrsim
0.03\,\text{mK}$).\cite{spr95,bun00a} The exchange coupling of liquid \He\ quasiparticles with
the localized \He\ spins, $J$, may modify the scattering of \He\
quasiparticles by the aerogel structure.\cite{spr96} Here we include the
effect of magnetic scattering of \He\ quasiparticles by polarizable \He\
spins coating the aerogel strands. The differential scattering of $\uparrow$
vs. $\downarrow$ spin quasiparticles by the polarized surface leads to an
additional contribution to the splitting of the $\uparrow\uparrow$ vs.
$\downarrow\downarrow$ transitions, $\lambda_{\text{\tiny J}}\propto J$,
which is determined by the non-magnetic, $u_0$, and exchange, $J$,
interactions and the density of \He\ coating the aerogel.\cite{bar00c} Below
we extend the analysis of Ref. \onlinecite{bar00c} and examine the role of
the exchange coupling on the possible $A_1-A_2$ splitting of the
superfluid phases of \He\ in aerogel.

%\section{Ginzburg-Landau Theory}

The suppression of the AB transition in both pure and disordered \He\
is quadratic in field on a scale set by $g_{\text{\tiny
AB}}\sim\text{mK/kG}^2$.\cite{ger01} Thus, for fields, $B\gg
B^*=\lambda^{\text{\tiny A}_1}/g_{\text{\tiny AB}}\sim 1\,\text{G}$,
and temperatures, $T\approx T_c$, the $\uparrow\downarrow$ pairs are
suppressed. In the field and temperature range of interest we can
restrict the full p-wave, spin-triplet order parameter to two
components, $\textsf{d}_+$ for $\uparrow\uparrow$ pairs and $\textsf{d}_-$ for
$\downarrow\downarrow$ pairs. We assume that the orbital state is the same for both
spin-components and of the axial (ABM) form,
$\chi(\hat\vp)=\hat\vp\cdot(\hat\vm+i\hat\vn)/\sqrt{2}$. Thus, in pure
\He\ the full Ginzburg-Landau (GL) free energy functional
(c.f. Ref. \onlinecite{vollhardt90}) reduces to\cite{mer73}
%----------------------------------------------------------------------------
\ber \Omega[\textsf{d}_+,\textsf{d}_-]    &=&\alpha\left(|\textsf{d}_+|^2 +|\textsf{d}_-|^2\right)
                   -\eta\,B\left(|\textsf{d}_+|^2 -|\textsf{d}_-|^2\right)
\nonumber\\
                   &+&\beta_{\text{\tiny 24}}\left(|\textsf{d}_+|^2 +|\textsf{d}_-|^2\right)^2
                   +4\beta_{\text{\tiny 5}}\,|\textsf{d}_+|^2\,|\textsf{d}_-|^2
\,,
\eer
%----------------------------------------------------------------------------
where $\alpha$, $\eta$ and $\beta_{\text{\tiny i}}$ are the known material parameters for
superfluid \He; $\alpha(T)=\tinyonethird N_f\ln(T/T_c)$ determines
the zero-field transition and $\eta=\tinyonethird N_f\lambda^{\text{\tiny A}_1}/T_c$
determines the $A_1$ ($\uparrow\uparrow$) transition, where $N_f$ is
the single-spin density of states at the Fermi level. The fourth-order
coefficients determine the relative stability of the possible phases. In
particular, $\beta_{\text{\tiny 24}}>0$, and $\beta_{\text{\tiny 5}}<0$
favors an equal-spin-pairing (ESP) phase with $|\textsf{d}_+|=|\textsf{d}_-|$. The linear
field term is symmetry breaking and competes with the fourth-order
terms. The latter wins at lower temperatures and gives rise to the $A_2$
transition where $\downarrow\downarrow$ spins condense,
with $T_c^{A_2}=T_c-\lambda^{\text{\tiny A}_2}\,B$ and
%----------------------------------------------------------------------------
\be
\lambda^{\text{\tiny A}_2} =
\left(\frac{\beta_{\text{\tiny 245}}}{-\beta_{\text{\tiny 5}}}\right)
\lambda^{\text{\tiny A}_1}
\,.
\ee
%----------------------------------------------------------------------------
Within the homogeneous, isotropic scattering model (HSM) the rotational symmetry of
\He\ in aerogel is preserved on the coherence length scale, and the GL free energy
has the same form as in pure \He, but with material parameters, $\bar\alpha$,
$\bar\eta$, etc., that are modified by the effects of scattering by the aerogel (we
use a `bar' to denote the material parameters in the presence of aerogel
scattering). These effects were calculated within the quasiclassical theory to
leading order in $T_c/E_f$ (weak-coupling), and one finds\cite{thu98}
%----------------------------------------------------------------------------
\begin{equation}
\bar\alpha= \onethird N_f\left[\ln(T/T_{c0}) - 2\,S_1(x)\right] \,,
\label{alpha}
\end{equation}
%----------------------------------------------------------------------------
where $x=v_f/2\pi T\ell$, and $\ell$ is the mean free path of quasiparticles
scattering off the aerogel. In Eq. (\ref{alpha}), and hereafter, we denote the
transition temperature for pure \He\ by $T_{c0}$. The superfluid transition in
aerogel is determined by the condition $\bar\alpha(T_c)=0$, and the function, $S_1$, is
%----------------------------------------------------------------------------
\be
S_1(x)=\sum_{n=0}^{\infty}\left(\frac{1}{2n+1+x}-\frac{1}{2n+1}\right)
\,.
\ee
%----------------------------------------------------------------------------
The parameter $\eta$ is directly proportional to the high-energy vertex,
$\Lambda$, and so is un-renormalized by impurities to leading order in
$\hbar/p_f\ell$. Thus, $\bar\eta=\eta=\tinyonethird
N_f\lambda^{\text{\tiny A}_1}/T_{c0}$, where $\lambda^{\text{\tiny
A}_1}$ is the rate for the splitting of the $A_1$-$A$ transition in pure \He.
Although $\eta$ is un-renormalized, the splitting parameters for the $A_1$ and $A_2$
transitions are renormalized by the impurity corrections
to transition temperature and, in general, the $\beta$ parameters.

The weak-coupling results for the fourth order coefficients are,\cite{thu98}
%----------------------------------------------------------------------------
\ber
\bar\beta_{\text{\tiny 24}}&=& -2\bar\beta_{\text{\tiny 5}}=
                              4(\bar\beta^{\text{\tiny wc}}+\bar\beta^{\bar\sigma})
\\
\bar\beta^{\text{\tiny wc}}    &=& \frac{N_f}{30\pi^2 T^2}\,S_3(x)
\nonumber \\
\bar\beta^{\bar\sigma}         &=& \frac{N_f}{9\pi^2 T^2}\,
                               (\bar\sigma_0-\onehalf)\,x\,S_4(x)
\,,
\eer
%----------------------------------------------------------------------------
where $\bar\sigma_0$ is the dimensionless, non-magnetic, s-wave scattering cross-section,
$0<\bar\sigma_0<1$ (see below), and
%----------------------------------------------------------------------------
\be
S_p(x)=\sum_{n=0}^{\infty}\left(\frac{1}{2n+1+x}\right)^p\,,\quad p>1
\,.
\ee
%----------------------------------------------------------------------------
Note that the ratio, $\bar r_{\beta}=\bar\beta_{245}/(-\bar\beta_5)\equiv 1$ in the
weak-coupling limit, even with (isotropic) impurity scattering. However,
this ratio deviates substantially from $1$ in pure \He, particularly at high
pressures, e.g. $r_{\beta}\approx 0.47$ at $p=33.4\,\text{bar}$.
Thus, the asymmetry of the $A_1-A$ vs. the $A_2-A$ transitions is a measure of
the strong-coupling corrections to the $\beta$ parameters,
$\delta\beta^{\text{\tiny sc}}=\beta-\beta^{\text{\tiny wc}}$, which are of order
%----------------------------------------------------------------------------
\be
\frac{\delta\beta^{\text{\tiny sc}}}{\beta^{\text{\tiny wc}}}
\sim\,\frac{T_c}{E_f}\,\langle |\Gamma_{\text{\tiny N}}|^2\rangle_{\text{\tiny FS}}
\,,
\ee
%----------------------------------------------------------------------------
compared to the weak-coupling values, where $\langle\ldots\rangle_{\text{\tiny FS}}$
is a Fermi-surface average of the normal-state quasiparticle-quasiparticle
collision rate, $\propto|\Gamma_{\text{\tiny N}}|^2$.

Corrections to the weak-coupling
$\beta$ parameters from quasiparticle scattering off the aerogel strands are of
order $\delta\bar\beta^{\text{\tiny wc}}/\beta^{\text{\tiny wc}}\sim x_c=v_f/2\pi\ell
T_c$, which is small for high-porosity aerogels, but comparable to the
strong-coupling corrections from quasiparticle-quasiparticle collisions. Based on the
suppression of $T_c$ and the aerogel mean free path we estimate $x_c\simeq 0.12$ at high
pressures.

If we neglect aerogel scattering corrections to
the intermediate quasiparticle states that enter the strong-coupling
self-energies,\cite{rai76} then the relative strong-coupling corrections for \He\ in
aerogel are scaled relative to their bulk ratios by the ratio of transition
temperatures,
%----------------------------------------------------------------------------
\be\label{sc_suppression_Tc}
\frac{\delta\bar\beta^{\text{\tiny sc}}}{\bar\beta^{\text{\tiny wc}}}=
\frac{\delta\beta^{\text{\tiny sc}}}{\beta^{\text{\tiny wc}}}\,
\left(\frac{T_c}{T_{c0}}\right)
\,.
\ee
%----------------------------------------------------------------------------
This approximation gives a good qualitative description of the suppression of
strong-coupling parameters for \He\ in aerogel as
measured by the field-dependence of the AB transition.\cite{ger01}

A theoretical calculation of impurity scattering corrections to the strong-coupling
$\beta$ parameters within the spin-fluctuation feedback theory of Brinkman and Anderson
was carried out by Baramidze and Kharadze.\cite{bar01} Their theory predicts a
suppression of strong-coupling effects with increased disorder, but at a rate that is
slower than that predicted just on the basis of the suppression of $T_c$. We can use the
results of Ref. \onlinecite{bar01} to estimate the strong-coupling correction to the
predicted $A_2$ transition for \He\ in aerogel. The results of Ref. \onlinecite{bar01}
depend on a high-energy vertex, which we determine by comparison with the magnitude of
strong-coupling ratio,
$r_{\beta}=\beta_{\text{\tiny 245}}/(-\beta_{\text{\tiny 5}})$ for pure \He. In particular,
we fix the ratio,
$\delta\beta^{\text{\tiny sc}}/\beta^{\text{\tiny wc}}$ for pure \He, using the measured
value of $r_{\beta}$:
$\tinyonehalf\delta\beta^{\text{\tiny sc}}/\beta^{\text{\tiny wc}}=
(1-r_{\beta})/(1+r_{\beta})$. Then, the impurity corrections to the strong-coupling
$\beta$-parameters calculated in Ref. \onlinecite{bar01},
give $\bar{r}_{\beta}=
(1-\tinyonehalf\,\delta\bar\beta^{\text{\tiny sc}}/\bar\beta^{\text{\tiny wc}})/
(1+\tinyonehalf\,\delta\bar\beta^{\text{\tiny sc}}/\bar\beta^{\text{\tiny wc}})$
where\footnote{This is the result for $\bar\sigma_0=1/2$. There is a correction to the
third term on the right-side of Eq. \ref{sc_suppression_sf} for $\bar\sigma_0\ne 1/2$.}
%---------------------------------------------------------------------------------
\begin{equation}\label{sc_suppression_sf}
\frac{\delta\bar\beta^{\text{\tiny sc}}}{\bar\beta^{\text{\tiny wc}}}=
\frac{\delta\beta^{\text{\tiny sc}}}{\beta^{\text{\tiny wc}}}
\left(\frac{T_c}{T_{c0}}\right)
\left(\frac{S_2(x_c)/S_2(0)}{S_3(x_c)/S_3(0)}\right)
\,.
\end{equation}
%---------------------------------------------------------------------------------
To leading order in the pair-breaking parameter,
$\delta\bar\beta^{\text{\tiny sc}}/\bar\beta^{\text{\tiny wc}}\approx
 (\delta\beta^{\text{\tiny sc}}/\beta^{\text{\tiny wc}})(1-a\,x_c)$. Based
 on Eq. (\ref{sc_suppression_Tc}), $a\simeq 2.47$. The rate of suppression
 is reduced to $a\simeq 1.28$ based on Eq. (\ref{sc_suppression_sf}). In what
 follows we use Eq. (\ref{sc_suppression_sf}) to estimate the suppression of the
 strong-coupling correction for $T_c^{\text{\tiny A}_2}$.
 This correction turns out to be small, and relatively unimportant on the scale of corrections
 that are required to explain the lack of an $A_1$-$A_2$ splitting for $B\le 5\text{kG}$.

%---------------------------------------------------------------------------------
\begin{figure}[t]
\includegraphics[width=8cm]{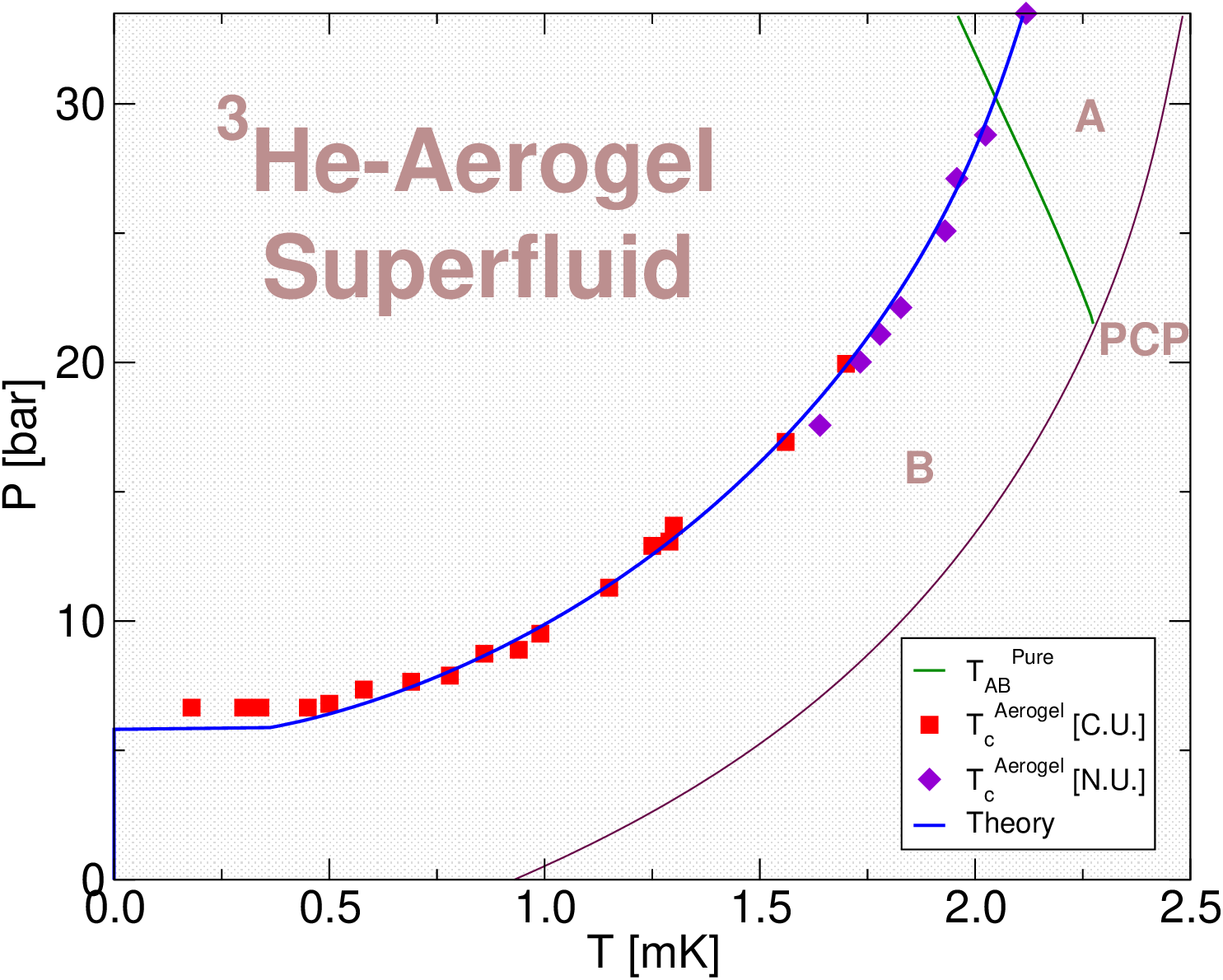}
         \caption{The phase diagram for \He\ in $98\,\%$ aerogel.
         The data are from Refs. \onlinecite{mat97} and \onlinecite{ger02c}.
         The theoretical curve is calculated from $\bar\alpha(T_c)=0$
         using Eq. (\ref{alpha}) in zero field
         with the effective pair-breaking parameter $\x$ evaluated
         with $\xi_a=502\,\text{\AA}$ and $\ell=1400\,\text{\AA}$. The phase boundaries for
         pure \He\ are shown for comparison.}
\label{fig-pvsT}
\end{figure}
%---------------------------------------------------------------------------------

Finally, before discussing the effects of the liquid-solid exchange coupling, we consider
a simplified version of an \textsl{inhomogeneous} scattering model discussed by
Ref. \onlinecite{thu98} that incorporates correlations of the aerogel. The length scale
at which aerogel reveals inhomogeneity, $\xi_a\sim 30-100\,\text{nm}$, is typically
comparable to the pair correlation length, $\xi$, and has a substantial effect on
the transition temperature of \He\ in aerogel, particularly at high pressures.
The inhomogeneity of the aerogel on scales $\xi_a\sim\xi$ leads to higher
superfluid transition temperatures than predicted by the HSM with the same
quasiparticle mean free path. Regions of lower aerogel density, of
size of order $\xi_a$, are available for formation of the condensate. Thus, the qualitative
picture is that of a random distribution of low density regions, `voids', with a typical
length scale, $\xi_a$, in an aerogel with a quasiparticle mean-free path, $\ell$.
When $\xi\sim\xi_a\ll\ell$, the superfluid transition is determined by
the pairbreaking effects of dense regions surrounding the `voids', and
scales as $\delta T_c/T_{c0}\propto -(\xi/\xi_a)^2$. However, when
the pair size is much larger than $\xi_a$ the aerogel is effectively homogeneous
on the scale of the pairs and pairbreaking results from homogeneous scattering
defined by the transport mean free path, which scales as
$\delta T_c/T_{c0}\propto -(\xi/\ell)$. This latter limit is achieved at low
pressures. We incorporate the correlation effect by introducing an effective
pairbreaking parameter in Eq. (\ref{alpha}) that interpolates between these two
limits, $x\to\x=x/(1+\zeta_a^2/x)$,
where $\zeta_a\equiv\xi_a/\ell$. This heuristic treatment of
aerogel correlations provides a good description of the pressure dependence
of $T_c$ in zero field for \He\ in aerogel over the whole pressure range, as shown in
Fig. \ref{fig-pvsT} for $\ell=1400\,\text{\AA}$ and $\xi_a=502\,\text{\AA}$.
Alternatively, we can adjust the mean-free path, $\ell$, with pressure in order
to simulate the correlation effect on $T_c$. However, we prefer to identify $\ell$ with
the pressure-independent geometric mean-free path and introduce aerogel correlation
effects via the effective pair-breaking parameter, $\tilde x$. In either scenario,
the GL theory for \He\ in aerogel predicts transitions for the
$A_1$ and $A_2$ phases, which correspond to the condensation of $\uparrow\uparrow$ and
$\downarrow\downarrow$ Cooper pairs as in pure \He; the transition
temperatures are of the same form,
%---------------------------------------------------------------------------------
\be
T_c^{\text{\tiny A}_1} = T_c + \bar\lambda^{\text{\tiny A}_1}\,B\,,\quad
T_c^{\text{\tiny A}_2} = T_c - \bar\lambda^{\text{\tiny A}_2}\,B
\label{Tc_A1&A2}
\,,
\ee
%---------------------------------------------------------------------------------
but with renormalized parameters,
%---------------------------------------------------------------------------------
\be
\hspace*{-2mm}
\bar\lambda^{\text{\tiny A}_1}=
    \lambda^{\text{\tiny A}_1}\left(\frac{T_c}{T_{c0}}\right)
    \left(1+2\x_c^{'}\,S_2(\x_c)\right)
\,,\,
\bar\lambda^{\text{\tiny A}_2}=\bar{r}_{\beta}\bar\lambda^{\text{\tiny A}_1}
\label{lambda_A1&A2}
\,,
\ee
%---------------------------------------------------------------------------------
where $\x_c'\equiv T_c d\x_c/dT_c$, and $\bar{r}_{\beta}=\bar\beta_{\text{\tiny
245}}/(-\bar\beta_{\text{\tiny 5}})$ is calculated including both impurity
scattering and strong-coupling corrections as described above. These results
predict an $A_1$-$A_2$ splitting of $\Del
T_c^{A_1-A_2}/B=\left(\bar\lambda^{\text{\tiny A}_1}+ \bar\lambda^{\text{\tiny
A}_2}\right) \simeq 6.3\mu\mbox{K/kG}$ at $p=33.4\,\text{bar}$, comparable
to that of pure \He. There is
currently no experimental evidence of an $A_1$-$A_2$ splitting in \He-aerogel.
Since the width of the transition is less than $20\,\mu\text{K}$, inhomogeneities
within the aerogel cannot account for the absence of the $A_1$-$A_2$ splitting.

For \He\ in aerogel an additional mechanism contributing to the splitting of
the $\uparrow\uparrow$ and $\downarrow\downarrow$ transitions is
possible.\cite{bar00c} It originates from an exchange coupling between liquid \He\
quasiparticles and the surface \He\ spins adsorbed on the silica
structure. Such a surface solid of \He\ has been observed for \He\
impregnated into silica aerogel. The signature is a Curie-like
susceptibility, $\chi_{\text{\tiny S}} = C/(T - \Theta_{\text{\tiny
S}})$, with a Curie temperature, $\Theta_{\text{\tiny S}}\approx
0.4\,\text{mK}$.\cite{spr96}

Thus, the model for scattering of quasiparticles by aerogel that we adopt
is a modified version of the scattering model described above which
includes an exchange coupling between \He\ quasiparticles in the liquid
and localized \He\ atoms bound to the silica aerogel structure. This coupling
is described by an exchange term in the quasiparticle-impurity potential,
%~~~~~~~~~~~~~~~~~~~~~~~~~~~~~~~~~~~~~~~~~~~~~~~~~~~~~~~~~~~~~~~~~~~~~~~~~~
\be\label{u_exchange}
u = u_0 + J\vS\cdot\vsigma
\,,
\ee
%~~~~~~~~~~~~~~~~~~~~~~~~~~~~~~~~~~~~~~~~~~~~~~~~~~~~~~~~~~~~~~~~~~~~~~~~~~
where $J$ is the liquid-solid exchange coupling, $\vS$ is the localized \He\ spin
operator and $\vsigma$ is the Pauli spin operator for the \He\ quasiparticles. There
are no direct measurements of $J$ for \He\ on aerogel, and theoretical calculations
for \He\ on planar substrates give indirect exchange interactions that vary over a wide range
of values, $J_{\text{\tiny ind}}\sim 0.1\,\mu\text{K}-1.0\,\text{mK}$, and may be either ferromagnetic
or anti-ferromagnetic depending on the specific mechanism and details of the theoretical
model (c.f. Ref. \onlinecite{god95}).

In a magnetic field, $\vB=-B\hat\vz$, the solid \He\ spins are polarized, $\vS=S(T,B)\hat\vz$,
with $S(T,B) = {\cal P}(B,T)\textsf{s}$, where $0\le {\cal P}\le 1$ is the fractional
polarization and $\textsf{s}=\tinyonehalf$. For sufficiently low fields, and
temperatures well above the ordering temperature for the solid \He\ spins, the polarization
is expected to be linear in field with ${\cal P}(B,T)\approx|\mu| B/k_B T$. In this
limit the  $A_1$-$A_2$ splitting
is given by Eqs. (\ref{Tc_A1&A2})-(\ref{lambda_A1&A2}), but
with $\bar\lambda^{\text{\tiny A}_1}\to\bar\lambda^{\text{\tiny A}_1}+\lambda_{\text{\tiny J}}$,
where $\lambda_{\text{\tiny J}}$ represents the effect of the surface polarization and
exchange coupling on the transition temperatures for $\uparrow\uparrow$ vs.
$\downarrow\downarrow$ pairs. The polarization-induced splitting,
$\lambda_{\text{\tiny J}}\propto J$, depends on the sign of the exchange coupling.
Thus, this term may either enhance or reduce the intrinsic splitting,
$\bar\lambda^{\text{\tiny A}_1}$. In what follows we calculate the exchange splitting,
$\lambda_{\text{\tiny J}}$, and discuss the result in relation to the
existing data for $T_c$.

%\section*{Model for $^3$He in Aerogel}

To calculate the liquid-solid exchange contribution to the $A_1$-$A_2$ splitting
we use the quasiclassical theory of superfluid \He,\cite{ser83} with effects of
scattering by the aerogel described by the HSM,\cite{thu98} modified  to include
the exchange coupling in Eq. (\ref{u_exchange}). The instability temperatures for
$\uparrow\uparrow$ and $\downarrow\downarrow$ Cooper pairs are obtained by solving
the weak-coupling gap equation for the spin-triplet components of the order
parameter,
%~~~~~~~~~~~~~~~~~~~~~~~~~~~~~~~~~~~~~~~~~~~~~~~~~~~~~~~~~~~~~~~~~~~~~~~~~~~~~~~
\ber\label{gap_equation}
\onethird\ln(T/T_{c0})\vDelta(\hat\vp) =
T\sum_{\varepsilon_n}\int\frac{d\Omega_{\hat\vp'}}{4\pi}(\hat\vp\cdot\hat\vp')\times
\nonumber\\
\left(\vf(\hat\vp';\varepsilon_n) - \pi\frac{\vDelta(\hat\vp')}{|\varepsilon_n|}\right)
\,,
\eer
%~~~~~~~~~~~~~~~~~~~~~~~~~~~~~~~~~~~~~~~~~~~~~~~~~~~~~~~~~~~~~~~~~~~~~~~~~~~~~~~
where $\vf(\hat\vp;\varepsilon_n)=(f_1,f_2,f_3)$ are the ``pair'' amplitudes for
the three spin-triplet states: $f_{\uparrow\downarrow}=f_3$,
$f_{\uparrow\uparrow}=(-f_1 + i f_2)$ and $f_{\downarrow\downarrow}=(f_1 + i
f_2)$. The pairing interaction and density of states at the Fermi level, as well
as the cutoff, have already been adsorbed into $T_{c0}$.

The scattering of quasiparticles off the aerogel structure is described by a random
distribution of scattering centers (``impurities''). The impurity self-energy,
to leading order in $\hbar/\tau E_f$, is determined by a t-matrix for multiple
scattering by a single impurity and the mean density of impurities,
%~~~~~~~~~~~~~~~~~~~~~~~~~~~~~~~~~~~~~~~~~~~~~~~~~~~~~~~~~~~~~~~~~~~~~~~~~~~~~~~
\be
\label{impurity-selfenergy}
\hat\Sig_{\mbox{\tiny imp}}(\hat\vp;\varepsilon_n)=
n_{\mbox{\tiny s}}\,\htm(\hat\vp,\hat\vp;\varepsilon_n)
\,.
\ee
%~~~~~~~~~~~~~~~~~~~~~~~~~~~~~~~~~~~~~~~~~~~~~~~~~~~~~~~~~~~~~~~~~~~~~~~~~~~~~~~

%\section*{Magnetic Scattering}
%
The model for scattering of quasiparticles by aerogel that we adopt
is described by an isotropic, non-magnetic scattering amplitude, $u_0$,
and an exchange term in the quasiparticle-impurity scattering potential;
in $4\times 4$ Nambu representation $\htu = u_0 \hat{1} + JS\,\hat\Sig_z$,
where $\hat{\vSig}=(\hat{\onematrix}+\hat\tau_{3})\vsigma/2 +
(\hat{\onematrix}-\hat\tau_{3})\vsigma^{\text{\tiny tr}}/2$ is the Nambu
representation for the quasiparticle spin.
For simplicity we also assume $J$ to be isotropic.
The t-matrix for repeated scattering of quasiparticles off a random distribution
of these polarized scattering centers is
%~~~~~~~~~~~~~~~~~~~~~~~~~~~~~~~~~~~~~~~~~~~~~~~~~~~~~~~~~~~~~~~~~~~~~~~~~~~~~~~
\be
\htm=\htu+N_f\htu\langle\hg\rangle\htm
\,,
\ee
%~~~~~~~~~~~~~~~~~~~~~~~~~~~~~~~~~~~~~~~~~~~~~~~~~~~~~~~~~~~~~~~~~~~~~~~~~~~~~~~
where $\langle\hg\rangle$ is the Fermi-surface-averaged propagator. For normal \He\ in aerogel
and even in the presence of magnetic fields and magnetic scattering,
the propagator reduces to $\hg_{\text{\tiny N}}=-i\pi\sgn{\epsilon_n}\hat\tau_3$.
Thus, the solution to the scattering t-matrix is given by
\be
\htm = \frac{1}{\pi N_f}
\left(\hat\onematrix + is_{\epsilon}\hat{\textsf{u}}\hat\tau_3\right)^{-1}\,
\hat{\textsf{\textsf{u}}}
\,,
\ee
where $s_{\epsilon}=\sgn{\epsilon_n}$, and the dimensionless scattering potential is
$\hat{\textsf{u}} = \textsf{u} \hat\onematrix + \textsf{v} \hSig_z$,
with $\textsf{u} = \pi N_f\,u_0$, $\textsf{v} = \pi N_f\,J\,S$.

For non-magnetic scattering ($S=0$) the t-matrix is
parameterized by the s-wave scattering phase shift, $\delta_0=\tan^{-1}(\textsf{u})$,
%~~~~~~~~~~~~~~~~~~~~~~~~~~~~~~~~~~~~~~~~~~~~~~~~~~~~~~~~~~~~~~~~~~~~~~~~~~~~~~~
\be
\htm = \frac{1}{\pi N_f}\,\sin\delta_0\,e^{-is_{\epsilon}\delta_0\hat\tau_3}
\,.
\ee
%~~~~~~~~~~~~~~~~~~~~~~~~~~~~~~~~~~~~~~~~~~~~~~~~~~~~~~~~~~~~~~~~~~~~~~~~~~~~~~~
In this minimal model for aerogel scattering, the
mean density of impurities and scattering rate for normal quasiparticles are fixed
by the mean free path, $\ell$, and scattering cross-section, $\sigma$,
%~~~~~~~~~~~~~~~~~~~~~~~~~~~~~~~~~~~~~~~~~~~~~~~~~~~~~~~~~~~~~~~~~~~~~~~~~~~~~~~
\be
n_{\text{\tiny s}} = \frac {1}{\sigma\ell}\,,
\quad\text{with}\quad
\sigma=\frac{4\pi\hbar^2}{p_f^2}
\,\bar\sigma_0
\,,
\ee
%~~~~~~~~~~~~~~~~~~~~~~~~~~~~~~~~~~~~~~~~~~~~~~~~~~~~~~~~~~~~~~~~~~~~~~~~~~~~~~~
where the normalized cross-section is related to the scattering potential by,
%~~~~~~~~~~~~~~~~~~~~~~~~~~~~~~~~~~~~~~~~~~~~~~~~~~~~~~~~~~~~~~~~~~~~~~~~~~~~~~~
\be
\bar\sigma_0 = \frac{\textsf{u}^2}{1 + \textsf{u}^2}
\,.
\ee
%~~~~~~~~~~~~~~~~~~~~~~~~~~~~~~~~~~~~~~~~~~~~~~~~~~~~~~~~~~~~~~~~~~~~~~~~~~~~~~~
Note that
$\bar\sigma_0\rightarrow 0$ is the Born scattering limit, while
$\bar\sigma_0\rightarrow 1$ is the unitary limit.

When $S\ne0$ there are different phase shifts for the scattering of
$\uparrow$ ($+$) and $\downarrow$ ($-$) spin quasiparticles, which we
parameterize as
\be
\delta^{\pm} = \delta_0 \pm \Del\delta
\,.
\ee
The t-matrix can now be expressed as,
\be
\begin{array}{lll}\label{t-matrix}
\htm &=& \frac{1}{\pi N_f}\,
\left\{
\sin\delta_0\cos(\Del\delta)\hat\onematrix +
\cos\delta_0\sin(\Del\delta)\hSig_z
\right\}\,
\\
&\times&e^{-is_{\epsilon}\delta_0\hat\tau_3}
\,e^{-is_{\epsilon}\Del\delta\hSig_z\hat\tau_3}
\,.
\end{array}
\ee
The quasiparticle-impurity scattering rates for $\uparrow$ and
$\downarrow$ quasiparticles are calculated from the retarded
self-energy, $\hSig^{\text{\tiny R}}_{\text{\tiny imp}} =
n_s\,\htm^{\text{\tiny R}}$, obtained from Eq. (\ref{t-matrix}) by setting
$s_{\epsilon}= +$. Thus, for quasi-particles the self-energy for spin
$\sigma_z=\uparrow$ and $\sigma_z=\downarrow$ becomes,
\be
\Sig^{\text{\tiny R}}_{\uparrow,\downarrow} = \Gamma_{\text{\tiny N}}\,
\sin\delta_{\uparrow,\downarrow}
\left(\cos\delta_{\uparrow,\downarrow} -i\sin\delta_{\uparrow,\downarrow}\right)
\,,
\ee
where $\Gamma_{\text{\tiny N}}=n_s/\pi N_f$. The scattering rates for
$\uparrow$ and $\downarrow$ spin quasiparticles are then,
\be
\frac{1}{2\tau_{\pm}} =
\Gamma_{\text{\tiny N}}\sin^2\delta^{\pm}=\Gamma_{\text{\tiny N}}\bar\sigma_{\pm}=
\Gamma_{\text{\tiny N}}\frac{(\textsf{u}\pm \textsf{v})^2}{1+(\textsf{u}\pm \textsf{v})^2}
\,,
\ee
where $\bar\sigma_{\pm}$ is the dimensionless cross-section for
scattering of $\uparrow$ vs. $\downarrow$ spin quasiparticles. In
both the unitary ($\delta_0\to\pi/2$) and the Born ($\delta_0\to 0$)
limits, the $\uparrow$ and $\downarrow$ spin scattering rates are
equivalent,
\begin{equation}
\frac{1}{2\tau_{\pm}}\rightarrow
\Bigg\{\begin{array}{l}
                    \Gamma_{\text{\tiny N}}\cos^2(\Del\delta)\,,\quad\delta_0=\pi/2 \\
                    \Gamma_{\text{\tiny N}}\sin^2(\Del\delta)\,,\quad\delta_0=0\,.
\end{array}
\end{equation}
Only when $\delta_0\ne 0,\pi$ is the scattering rate for
$\uparrow$ and $\downarrow$ spin quasiparticles different.
In general, we can parameterize the scattering rates as
\be
\frac{1}{\tau^{\pm}} =
\frac{1}{\bar\tau}\pm\frac{1}{\tau_{\text{\tiny S}}}
\,,
\ee
or equivalently,
\be
\frac{1}{\tau_{\text{\tiny S}}}=\frac{1}{\bar\tau}\,
\left(\frac{\bar\sigma_{+}-\bar\sigma_{-}}{\bar\sigma_{+}+\bar\sigma_{-}}\right)
\,,
\ee
where $1/\bar\tau$ is the polarization-independent scattering rate.
It is convenient to express the normal-state self-energy in terms of base
particle-hole matrices,
\be\label{Sig_N}
\hSig_{\text{\tiny N}}=\Sig_{11}\hat\onematrix
     +\Sig_{13}\hSig_z
     +\Sig_{31}\hat\tau_3
     +\Sig_{33}\hat\tau_3\hSig_z
\,,
\ee
with components
\ber
\Sig_{11}&=&+\onehalf\Gamma_{\text{\tiny N}}\sin(2\delta_0)\cos(\Del\delta)\,, \\
\Sig_{13}&=&+\onehalf\Gamma_{\text{\tiny N}}\cos(2\delta_0)\sin(\Del\delta)\,, \\
\Sig_{31}&=&-\frac{i}{2}\Gamma_{\text{\tiny N}}\,s_{\epsilon}[1-\cos(2\delta_0)\cos(2\Del\delta)]\,,\\
\Sig_{33}&=&-\frac{i}{2}\Gamma_{\text{\tiny N}}\,s_{\epsilon}\sin(2\delta_0)\sin(2\Del\delta)\,.
\eer

%\section*{A-A$_1$ Splitting}

To calculate the instability temperatures for $\uparrow\uparrow$ and
$\downarrow\downarrow$ pairs we need the off-diagonal propagator to linear
order in the pairing self-energy. Thus, we expand the transport equation,
self-energies and normalization condition in powers of $\hat\Del$. The
zeroth-order terms are the normal-state propagator and self-energy (Eq.
(\ref{Sig_N})). To first-order we obtain,\cite{ser83}
%------------------------------------------------------------------------------
\be
\left[i\varepsilon_n\hat\tau_3 - \hSig_{\text{\tiny N}}\,,\,\hg^{(1)}\right] =
\left[\hDel\,,\,\hg_{\text{\tiny N}}\right]
\,,
\ee
%------------------------------------------------------------------------------
and $\hat\tau_3\hg^{(1)} + \hg^{(1)}\hat\tau_3 = 0$
from the normalization condition.
We reduce the equations to $2\times 2$ spin-space by writing
$\hat{\textsf{H}}_{\text{\tiny N}}=i\epsilon_n\hat\tau_3-\hSig_{\text{\tiny N}}=
\tinyonehalf(\hat\onematrix+\hat\tau_3)\textsf{H}_\text{{\tiny N}}+
\tinyonehalf(\hat\onematrix-\hat\tau_3)\bar{\textsf{H}}_\text{{\tiny N}}$
with $\textsf{H}_\text{{\tiny N}}=i\epsilon_n-\Sig_{\text{\tiny N}}$
and $\bar{\textsf{H}}_\text{{\tiny N}}=-i\epsilon_n - \bar\Sig_{\text{\tiny N}}$.
Note that $\hg^{(1)}$ is purely off-diagonal, with the upper-right pair amplitude
satisfying the equation in spin-space,
%------------------------------------------------------------------------------
\be
\textsf{H}_{\text{\tiny N}}\,f^{(1)}-f^{(1)}\,\bar{\textsf{H}}_{\text{\tiny N}}
= 2i\pi\,\sgn{\epsilon_n}\,\Delta
\,.
\ee
%------------------------------------------------------------------------------
Projecting out the spin-triplet components, we obtain,
$\left(i\epsilon_n-\Sig_{\pm}\right)\,f_{\pm}=i\pi\,\sgn{\epsilon_n}\,\Delta_{\pm}$
with $\Sig_{\pm}=\Sig_{31}\pm\Sig_{33}=-i\sgn{\epsilon_n}/2\tau_{\pm}$, and for
$f_{\uparrow\uparrow(\downarrow\downarrow)}\equiv f_{+(-)}=\mp f_1 + i f_2$,
%------------------------------------------------------------------------------
\be
\label{lge_up}
f_{\pm}=\frac{\pi\Delta_{\pm}}{|\epsilon_n|+1/2\tau_{\pm}}
\,.
\ee
%------------------------------------------------------------------------------
The linearized gap equations for $\Delta_{\pm}$ are given by Eq. (\ref{gap_equation})
with $\vf\rightarrow f^{\pm}$, $\vDelta\rightarrow\Delta^{\pm}$ and $T\rightarrow T_c^{\pm}$.
For non-unitary, axial states,
%------------------------------------------------------------------------------
\be
\Delta_{\pm}(\hat\vp)=\textsf{d}_{\pm}\,\left(\hat{p}_x+i\hat{p}_y\right)/\sqrt{2}
\,,
\ee
%------------------------------------------------------------------------------
the eigenvalue equation for $\textsf{d}_{\pm}$ yields the weak-coupling equation for the
instability temperatures, $T_c^{\pm}$. In the absence of the polarization, the aerogel
transition temperature is given by
%------------------------------------------------------------------------------
\be\label{Tc}
\ln\left(T_c/T_{c0}\right)=2\,S_1(x_c)
\,,
\ee
%------------------------------------------------------------------------------
where $x_c=1/2\pi\tau T_c$, and the spin-independent rate for quasiparticles scattering
off the aerogel is given by
%------------------------------------------------------------------------------
\be
\frac{1}{2\tau}=\Gamma_{\text{\tiny N}}\,\sin^2\delta_0\equiv
                \Gamma_{\text{\tiny N}}\,\bar\sigma_0
\,.
\ee
%------------------------------------------------------------------------------

In the presence of a liquid-solid exchange coupling, and polarization of the
solid \He, the instability temperatures for the $\uparrow\uparrow$ and
$\downarrow\downarrow$ condensates are given by
%------------------------------------------------------------------------------
\be\label{Tc_+/-}
\ln\left(T_c^{\pm}/T_{c0}\right)=2\,S_1(x^{\pm})
\,,
\ee
%------------------------------------------------------------------------------
where $x^{\pm}\equiv 1/2\pi\tau_{\pm}T_c^{\pm}$. For $u_0\ne 0$, the leading
order polarization correction to the scattering cross-sections gives
%------------------------------------------------------------------------------
\be
\frac{1}{\bar\tau}=\frac{1}{\tau}
\,,\quad
\frac{1}{2\tau_{\text{\tiny S}}} =
2 n_s J S(T,B)\,\sqrt{\bar\sigma_0}(1-\bar\sigma_0)^{3/2}
\,.
\ee
%------------------------------------------------------------------------------
In the low-field region, and above the magnetic ordering temperature,
${\cal P}=|\mu| B/k_B T_c$, and we obtain,
%------------------------------------------------------------------------------
\be\label{lambda_J}
\hspace*{-2mm}
\lambda_{\text{\tiny J}}=g_{\text{\tiny J}}\,\left(\frac{|\mu|}{k_B}\right)\,
\left(\frac{(1-\bar\sigma_0)^{3/2}}{\sqrt{\bar\sigma_0}}\right)
\left(\frac{-2x_c\,S_2(x_c)}{1-2x_c\,S_2(x_c)}\right)
\,,
\ee
%------------------------------------------------------------------------------
where $x_c = (\xi_0/\ell)(T_{c0}/T_c)$, and the dimensionless exchange coupling is
%------------------------------------------------------------------------------
\be
g_{\text{\tiny J}}=2\pi N_f\,J\,\ess
\,.
\ee
%------------------------------------------------------------------------------
Note the impurity-induced exchange splitting vanishes in the unitary
limit.\footnote{The splitting also vanishes in the Born limit; although Eq.
(\ref{lambda_J}) is not valid in the Born limit since it is based on a expansion
of $J/u_0$. Nevertheless, in the limit $u_0=0$, i.e. with only pure exchange
coupling the cross-sections for $\uparrow$ and $\downarrow$ spin scattering are
equal and the splitting vanishes.} Equation (\ref{lambda_J}) is easily generalized
to include aerogel correlations within the heuristic `random void' model described
above; the result for $\lambda_{\text{\tiny J}}$ has the same form
as Eq. (\ref{lambda_J}), but with
$-x_c\,S_2(x_c)\rightarrow\tilde{x}_c'\,S_2(\tilde{x}_c)$, where
$\tilde{x}_c=x_c^2/(x_c+\zeta_a^2)$, $x_c=1/2\pi\tau T_c$, and
$\tilde{x}_c'\equiv T_c\,d\tilde{x}_c/dT_c$.

%----------------------------------------------------------------------------------------
\begin{figure}[t]
\includegraphics[width=8.5cm]{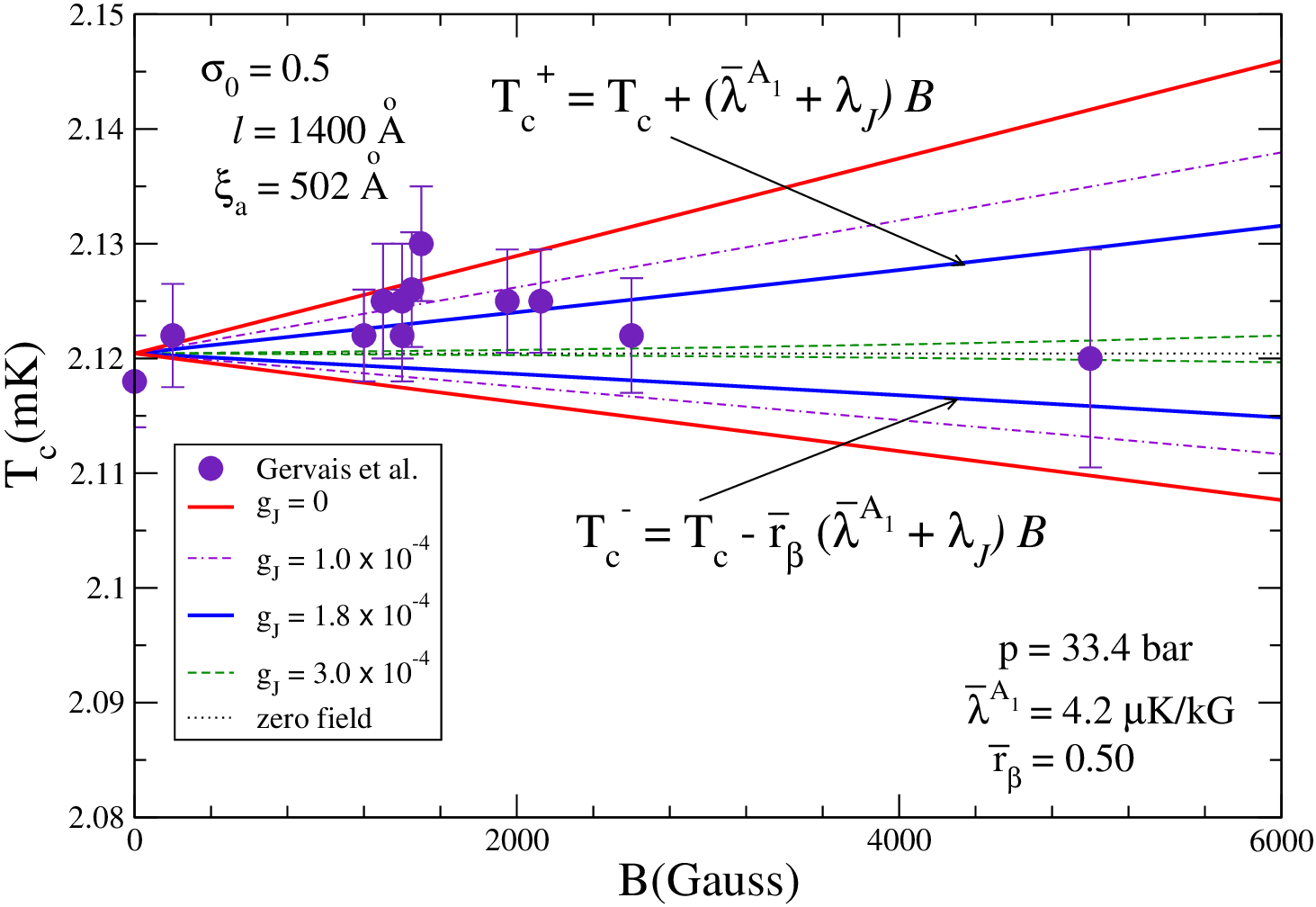}
         \caption{The low-field (linear) splitting of $T_c$ with magnetic
         field for \He\ in aerogel with a mean free path of
         $\ell=140\mbox{nm}$, a correlation length of $\xi_a=50\text{nm}$
         and a typical cross-section, $\bar\sigma_0=1/2$. The
         splitting for \He-aerogel without liquid-solid exchange
         is indicated by the solid (red) lines. The data points
         are taken from Ref. \onlinecite{ger02c}.}
\label{fig:Tc_split_Compare}
\end{figure}
%----------------------------------------------------------------------------------------

The effects of the liquid-solid exchange coupling, $g_{\text{\tiny J}}$,
and the polarization of the solid \He\ coating the aerogel strands on the
$A_1$-$A_2$ splitting are shown in Fig. \ref{fig:Tc_split_Compare}, and
compared with measurements of the superfluid transition in 98\% aerogel
reported in Ref. \onlinecite{ger02c}; these authors found no evidence of
an $A_1$-$A_2$ splitting for fields up to $B=5\,\mbox{kG}$. The data for
the superfluid transition of \He\ in 98\% aerogel for fields up to
$B=5\,\text{kG}$ are shown in Fig. \ref{fig:Tc_split_Compare}. The error
bars are conservative estimates of the uncertainty in defining $T_c$; the
experiment shows no evidence of a splitting to within the error of
determining $T_c$, and consequently we can assume that the splitting to be
less than the error bars for $T_c$.\footnote{W. P. Halperin, private
communication.}

The calculation of the $A_1$-$A_2$ splitting includes aerogel correlations, which are
most important at high pressures. Both the mean-free path, $\ell$, and the aerogel
correlation length, $\xi_a$, contribute. The values of $\ell=1400\,\text{\AA}$ and
$\xi_a=502\,\text{\AA}$ correspond to $T_c=2.12\,\text{mK}$ at $p=33.4\,\text{bar}$
and yield close agreement with $T_c(p)$ over the full pressure range. The
dimensionless cross-section, $\bar\sigma_0$, is not known with any certainty; there
is likely a distribution of cross-sections provided by the aerogel. In the absence of
detailed knowledge we assume an average value of $\bar\sigma_0=1/2$. The values of
$\lambda^{\text{\tiny A}_1}$ and $\lambda^{\text{\tiny A}_2}$ for pure \He, and thus
the strong-coupling parameter, $r_{\beta}$, are taken from Ref. \onlinecite{isr84}.
The effects of non-magnetic scattering by aerogel lead to small corrections for
$\bar\lambda^{\text{\tiny A}_1}$ and $\bar\lambda^{\text{\tiny A}_2}$; these terms
alone (shown in Fig. \ref{fig:Tc_split_Compare} as $g_{\text{\tiny J}}=0$) generate
an $A_1$-$A_2$ splitting that is substantially larger ($\approx\times 2$) than the
error reported for the superfluid \He\ transition in Ref. \onlinecite{ger02}. An
anti-ferromagnetic exchange coupling ($g_{\text{\tiny J}}>0$) decreases the
$A_1$-$A_2$ splitting. The magnitude of the predicted splitting is reduced to lie
within the error bars for $T_c$ for $g_{\text{\tiny J}}=1.8\times 10^{-4}$, which
corresponds to an exchange coupling of $J\simeq 0.1\,\text{mK}$ per liquid \He\ spin.

%----------------------------------------------------------------------------------------
\begin{figure}[t]
\includegraphics[width=8.5cm]{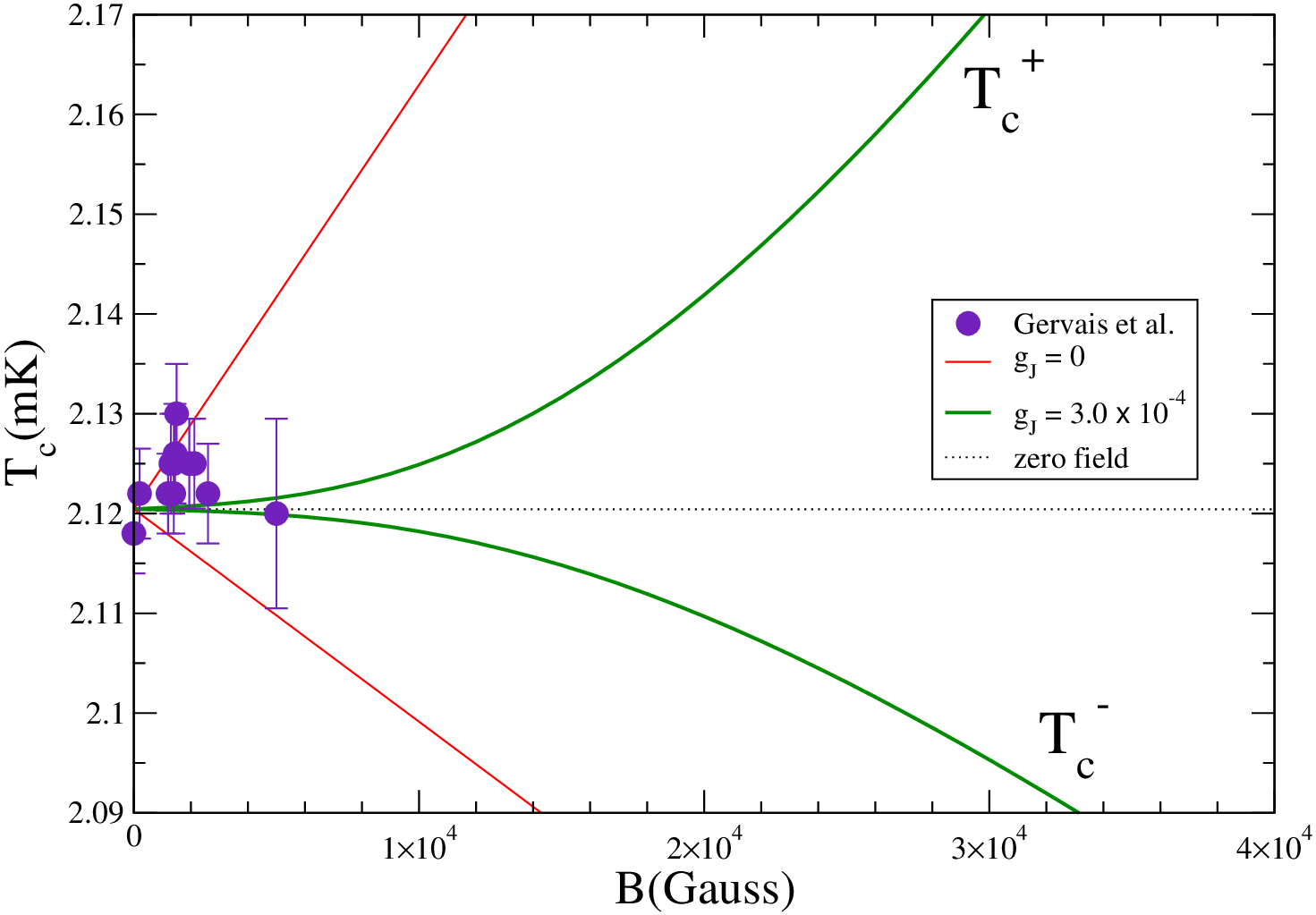}
         \caption{The field-evolution of the splitting of $T_c$
         for \He\ in aerogel with the same parameters as those used
         in Fig. \ref{fig:Tc_split_Compare}. The linear field splitting
         expected in the absence of polarized solid \He\
         is indicated by the solid (red) lines. The
         nonlinear field evolution of the splitting
         is indicated by the thick (green) lines and corresponds to
         the value of $g_J=3.0\times 10^{-4}$. The Curie temperature is
         taken from Ref. \onlinecite{spr95},
         $\Theta_{\text{\tiny S}}\simeq 0.4\,\text{mK}$,
         and the exchange field is,
         $B_{\text{\tiny S}}=k_B\Theta_{\text{\tiny S}}/|\mu|\simeq 5.14\,\text{kG}$.
         The data points are from Ref. \onlinecite{ger02c}.}
\label{fig:Tc_split_nonlinear}
\end{figure}
%----------------------------------------------------------------------------------------

The existing data, while suggestive that the liquid-solid layer coupling may be playing
an important role in suppressing the $A_1$-$A_2$ splitting, is not conclusive. If
scattering by polarized \He\ is responsible for the suppressed $A_1$-$A_2$ splitting for
pure \He\ in aerogel, then heat capacity or acoustic attenuation measurements with
\Hefour\ added to displace the solid \He, should exhibit an $A_1$-$A_2$ splitting that is
comparable to that of pure bulk \He. Measurements of Sprague et al.\cite{spr96} at
$p=18.7\,\text{bar}$ and at $B=1.47\,\text{kG}$ do show and increase in $T_c$ from
$1.69\,\text{mK}$ without \Hefour\ coverage, to $1.76\,\text{mK}$ with the
addition of one monolayer of \Hefour\ to remove the solid \He; thus, $\Del T_c \simeq
70\,\mu\text{K}$. By comparison, if we suppress the polarization component of the
scattering rate in our theoretical calculation we obtain and increase in $T_c$ from
the conventional component of the
$A_1$-$A_2$ splitting of $\Del T_c^{\text{\tiny A}_1-\text{\tiny A}_2}=
3.1\,\mu\text{K}/\text{kG}\,B\simeq 4.6\mu\text{K}$, which is more than an order of
magnitude smaller than the change in $T_c$ observed by adding \Hefour. Thus,
the addition of \Hefour\ also modifies the non-magnetic contribution to the pairbreaking,
and this effect is dominant at these low fields.

Measurements on pure \He\ in aerogel at higher fields should not suffer from this
problem and should be able to resolve some or all of the uncertainty in the
mechanism suppressing the $A_1$-$A_2$ splitting at low fields.
In particular, if an exchange coupling, $J\approx 0.1-0.2\,\text{mK}$, is responsible for the
suppressed $A_1$-$A_2$ splitting at $B\le 5\,\text{kG}$, then for higher fields,
$B\gg B_{\text{\tiny S}}= k_B\Theta_{\text{\tiny S}}/|\mu|\approx 5\,\text{kG}$ the
polarization of the solid \He\ should saturate, producing a field-independent shift
from scattering off the polarized \He, and an $A_1$-$A_2$ splitting that increases with
field, for $B\gg B_{\text{\tiny S}}$, at a rate comparable to that for pure \He.

The Curie temperature for the solid \He\ provides the temperature and field scale for the polarization, i.e.
${\cal P}(B/B_{\text{\tiny S}},T/\Theta_{\text{\tiny S}})$. In order to estimate the field-dependence of the
$A_1$-$A_2$ splitting at higher fields we use the mean-field theory for the $\textsf{s}=1/2$, near-neighbor
Heisenberg ferromagnet to calculate the polarization.\cite{ashcroft75} The result is shown in Fig.
\ref{fig:Tc_split_nonlinear} for the same parameters used to obtain the low-field suppression of the
$A_1$-$A_2$ splitting shown in Fig. \ref{fig:Tc_split_Compare}. Thus, even if fully suppressed at low fields,
$B\lesssim B_{\text{\tiny S}}$, the $A_1$-$A_2$ splitting should emerge for fields above $B\approx
20\text{kG}$.

\medskip

We thank Yuriy Bunkov, Henri Godfrin, and Bill Halperin for useful discussions, and
acknowledge support from NSF grant DMR-9972087.

\end{document}